\documentclass[12pt,preprint]{aastex}

\slugcomment{Accepted for Publication in Astrophysical Journal Letters}

\shorttitle{Type III Radio Storm}
\shortauthors{Eastwood et al.}

\begin{document}


\title{On the Brightness and Waiting-time Distributions of a Type III Radio Storm observed by STEREO/WAVES}


\author{J. P. Eastwood, \altaffilmark{1} M. S. Wheatland, \altaffilmark{2} H. S. Hudson, \altaffilmark{1}  S. Krucker, \altaffilmark{1} S. D. Bale, \altaffilmark{1,3} M. Maksimovic, \altaffilmark{4} K. Goetz, \altaffilmark{5} and J.-L. Bougeret\altaffilmark{4}}
\altaffiltext{1}{Space Sciences Laboratory, University of California, Berkeley, CA 94720, USA}
\altaffiltext{2}{Sydney Institute for Astronomy, School of Physics, The University of Sydney, NSW 2006, Australia}
\altaffiltext{3}{Department of Physics, University of California, Berkeley, CA 94720, USA}
\altaffiltext{4}{LESIA, Observatoire de Paris, F-92195 Meudon, France}
\altaffiltext{5}{School of Physics and Astronomy, University of Minnesota, Minneapolis, MN 55455, USA}

\email{eastwood@ssl.berkeley.edu}



\begin{abstract}
Type III solar radio storms, observed at frequencies below $\sim$ 16 MHz by space borne radio experiments, correspond to the quasi-continuous, bursty emission of electron beams onto open field lines above active regions. The mechanisms by which a storm can persist in some cases for more than a solar rotation whilst exhibiting considerable radio activity are poorly understood. To address this issue, the statistical properties of a type III storm observed by the STEREO/WAVES radio experiment are presented, examining both the brightness distribution and (for the first time) the waiting-time distribution. Single power law behavior is observed in the number distribution as a function of brightness; the power law index is $\sim$ 2.1 and is largely independent of frequency. The waiting-time distribution is found to be consistent with a piecewise-constant Poisson process. This indicates that during the storm individual type III bursts occur independently and suggests that the storm dynamics are consistent with avalanche type behavior in the underlying active region.
\end{abstract}






\section{Introduction}

Type III solar radio storms consist of many individual type III radio bursts produced quasi-continuously over several days \citep{fain70a,fain70b,fain71}. Such storms can persist for long intervals -- they have been observed for up to half a solar rotation, and in some cases over a whole rotation  \citep{fain70a}. It has been shown that they are associated with both active regions and metric type I storms \citep{kays87,gopa04}. The type III storm emission is produced by electron beams on open field lines \citep{boug84a, boug84b}, and so type III storm activity may shed light on the dynamics of the open/closed field line boundaries associated with the underlying active region. The dynamics of type III storms are still not well understood, particularly the mechanisms by which they are able to persist for such long times whilst at the same time generating considerable numbers of radio bursts. Furthermore, type III storms have also been observed as precursors to coronal mass ejections \citep{rein01}, and so a better understanding of their properties may help to elucidate the dynamics of active regions prior to eruptive events.

One important statistical property is the distribution of power $S$ in the individual bursts. In their initial type III storm study, \citet{fain70a} noted that there were more faint bursts than bright ones. More recently, \citet{mori07} studied the statistics of type III bursts using Geotail (24Hz - 800 kHz) and Akebono (20kHz - 5MHz) radio data. Using observations covering 4 Bartels rotations and containing both ordinary type III emission and intervals of type III storm activity, they found that the overall occurrence of type III bursts followed a double power law; for example, at a frequency of 860 kHz and below an intensity of $\sim 10^{-15}$ V$^{2}$m$^{2}$Hz$^{-1}$ ($\sim$ 2.7 $\times$ 10$^{-18}$ Wm$^{-2}$Hz$^{-1}$), the number distribution (simply the number of bursts at a given brightness) followed a $S^{-3.6}$ power law whereas above this limit, a $S^{-0.52}$ power law was observed. \citet{mori07} named the lower flux distribution `micro-type III bursts', and concluded that although they were associated with the same active regions as the `ordinary' type III bursts, they were produced by independent, coexisting processes. This is consistent with the association of the Type III storms with meter-wave Type I `noise storms', which do not have the close association with flares that characterizes ordinary type III bursts \citep[e.g.][]{gopa04}. In comparison, \citet{fitz76} found in a study of ordinary type III bursts observed over a period of three months that the number/unit flux density fell off with increasing flux following a $S^{-\beta}$ power law with $\beta$ varying from 1.33 at 110 kHz to 1.69 at 4.9 MHz. This in fact appears compatible with the \citet{mori07} study, since a $S^{-1.69}$ power law in the number/unit flux density corresponds to a $S^{-0.69}$ power law in the number distribution. For comparison, at higher frequencies (164 MHz, 237 MHz and 327 MHz), the number distribution of type I bursts in radio noise storms has been found to follow a similarly steep $S^{-3}$ power law \citep{merc97}.

A second statistical property is the Waiting-Time Distribution (WTD), that is the distribution of times between individual type III events within the storm. WTD analysis has been applied to soft X-ray flares \citep{whea00, whea02} and Coronal Mass Ejections \citep{whea03} and in both cases it has been shown that the data are consistent with a time-dependent Poission process, in which events occur independently. The WTD provides models with extra constraints; for example, flare avalanche models \citep[e.g.][]{lu1993} predict that flares occur independently, and so the WTD is an important observable. 

To our knowledge the WTD has not been used to examine type III storms, particularly at the low radio frequencies typically measured by spacecraft (below $\sim$ 16 MHz). In this Letter we analyze both the brightness distribution and the WTD of a type III storm observed by the S/WAVES radio experiment \citep{boug08} on board the STEREO spacecraft \citep{kais08}, in order to examine in more detail the underlying processes responsible for type III storm emission.

\section{Observations}

Each STEREO spacecraft is equipped with an identical S/WAVES instrument designed to measure both in-situ plasma waves and freely propagating radio emission in the frequency range 2.5 kHz - 16.025 MHz \citep{boug08}. S/WAVES uses three orthogonal monopole electric antennas \citep{bale08} connected to two instrument channels; in a typical operating mode, one channel is connected to a pseudo dipole and the other to a monopole. Within each channel, there is a high frequency receiver (HFR) and a low frequency receiver (LFR). The observations presented here were obtained shortly after launch using the HFR which covers the frequency range 125 kHz - 16.025 MHz, before the lunar flybys that sent the STEREO spacecraft into their final heliocentric orbits and before many of the other instruments became operational. Consequently the radio data from the two spacecraft are essentially identical, and so here we present the analysis from STEREO A.

\begin{figure}
\includegraphics[scale=1]{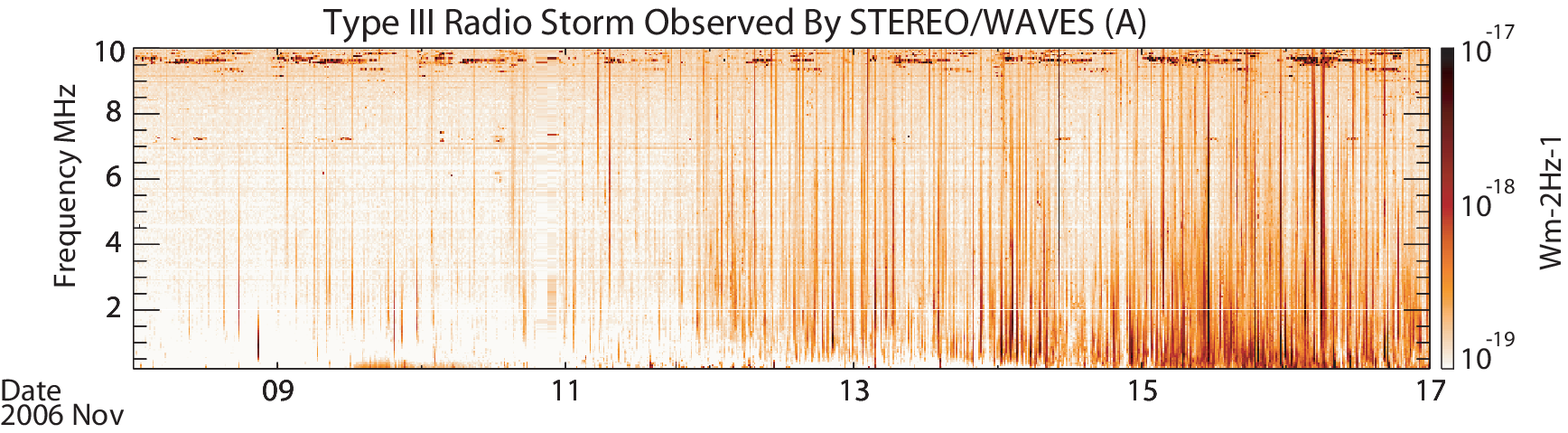}
\caption{\label{fig:fig1}STEREO-A S/WAVES observations of radio emission from 8 Nov 2006 0000UT to 17 Nov 2006 0000UT (9 days). A type III storm associated with Active Region 10923 was observed, beginning on 11 Nov. Terrestrial RFI is observed periodically between 9.5 MHz and 10 MHz but is avoided in the analysis. The white line at 2 MHz is an artifact of the plotting software.}
\end{figure}

Figure~\ref{fig:fig1} shows measurements from the S/WAVES experiment on STEREO A measured between 0.2 MHz and 10 MHz from 8 November 2006 0000 UT - 17 November 2006 0000 UT; 9 days of data are shown. The data are from channel 1 (corresponding to the $E_{x}$ - $E_{y}$ dipole), and have been calibrated in units of incident flux $S$ [Wm$^{-2}$Hz$^{-1}$] following the procedure described by \citet{east09}. The calibration procedure is most accurate below $\sim$ 10 MHz; at higher frequencies, the antenna resonance is less well modeled. Lines of Radio Frequency Interference (RFI) at fixed frequencies have been removed by averaging across the frequency band. Terrestrial RFI is still observed periodically between 9.5 MHz and 10 MHz; in the following analysis, frequencies have been chosen to avoid this interference. 

Type III emission began to be observed intermittently as early as 9 November 2006, and the storm itself was established by 11 November 2006. Active region 10923 was the only active region on the disk during this time and so we associate it with the type III storm \citep{kays87} although it appears that the storm developed after the active region appeared on the disk. Over the next 6 days, the number, frequency and intensity of type III emissions increased. On 17 November, the STEREO spacecraft executed an Earth perigee pass. Whilst the storm continued, S/WAVES observed overwhelming terrestrial RFI at all frequencies. Active region 10923 was close to the limb by this time, and a drop in the number and intensity of type III bursts on 18 November was followed by the leading spot of AR 10923 rotating past the limb on 20 November. This confirmed the association of the active region with the type III bursts \citep{boug84b}, but suggests that the activity is more associated with the leading edge of the region. In this Letter the interval 11 November 2006 0000 UT - 17 November 2006 0000 UT has been analyzed.

\section{Analysis and Results}

To examine the properties of this storm interval in more detail, time series of data at specific frequencies were extracted and analyzed. Identification of individual type III bursts was made in the following way. We first established a threshold brightness $S_0$ and defined a burst as an interval during which time the brightness $S$ satisfies $S > S_0$. The burst was then assigned a brightness $S_{\rm{burst}}$ = $S_{\rm{max}} - S_{\rm{g}}$, the maximum brightness during the burst interval less the galaxy brightness $S_{\rm{g}}$ \citep{cane79}. The time of the burst is similarly defined as the time the maximum brightness was observed. Inspection of the data during the most active intervals shows that in general, bursts are typically observed to last only one timestep ($\sim$ 25 s), and bursts lasting multiple time steps are rare. Accordingly mis-counting from event overlap is not significant. This results in a set of $N$ bursts and from their observed times, a set of $N - 1$ waiting times. 

We first consider the brightness distribution $f(S)$, which we define as the number of bursts per unit brightness. If this distribution follows a power law

\begin{equation}
\label{eq1}
f(S) = AS^{-\delta}
\end{equation}

\begin{flushleft}
then  the maximum likelihood method \citep{bai93} can be used to show that the most likely value of the power law index $\delta_{m}$ is given by 
\end{flushleft}

\begin{equation}
\label{eq2}
\delta_{m} = \frac{N}{\sum^{N}_{i=1}\ln(S_{i}/S_{0})}+1,
\end{equation}

\begin{flushleft}
where $S_i$ is the brightness of the $i^{th}$ burst. The error associated with $\delta_{m}$ is given by \citep{whea04}:
\end{flushleft}

\begin{equation}
\label{eq3}
\sigma_{\delta} \sim (\delta_{m} - 1) N^{-1/2}.
\end{equation}

Rather than examining the brightness distribution directly, it is preferable to plot the cumulative brightness distribution, which avoids binning of the data, and more easily shows the effect of the choice of $S_0$. The cumulative distribution also follows a power law as shown by equation (\ref{eq4}):

\begin{equation}
\label{eq4}
N_{S > S_0} = \int^{\infty}_{S_0} AS^{-\delta} dS = \frac{A}{\delta - 1} S_0^{-\delta + 1}.
\end{equation}

\begin{figure}
\includegraphics[scale=1]{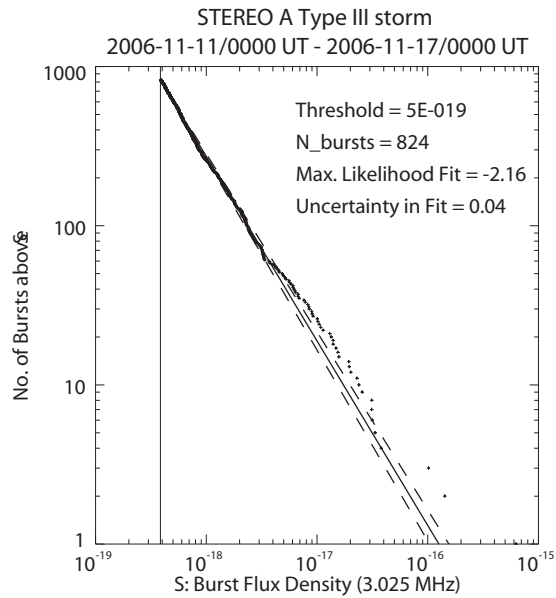}
\caption{\label{fig:fig2}Cumulative distribution of burst brightness at 3.025 MHz, together with the maximum likelihood fit.}
\end{figure}

This analysis was applied to the 3.025 MHz time series. Figure~\ref{fig:fig2} shows the distribution of cumulative burst brightness (points). A threshold $S_0 = 5 \times 10^{-19}$ Wm$^{-2}$Hz$^{-1}$ is used, well above the average galactic brightness of $\sim 1.1 \times 10^{-19} $ Wm$^{-2}$Hz$^{-1}$. This results in a set of 824 bursts. In this case, the power law index associated with the brightness distribution is $\delta = 2.16 \pm 0.04$. The power law derived from the data using the maximum likelihood method is shown as a solid line in Figure~\ref{fig:fig2}; dashed lines show the power law plus/minus the error calculated using equation (\ref{eq3}). The data in general follow the power law and the significance of any departure in the final few points is marginal; there is no substantial evidence for a break to a different power law index although a detailed quantitative model comparison has not been done. If a higher value of $S_0$ is chosen, this has the effect of removing points on the left hand side of the cumulative distribution, but does not change the index of the power law. If a smaller value of $S_0$ is chosen, the data eventually rolls over to the background level. This procedure was repeated for other frequencies and the results are shown in Table~\ref{table1}. It can be seen that there is very little variation in the power law index as the frequency changes. A larger power law index is seen compared to the study of type III bursts \citep{fitz76}.

\begin{figure}
\includegraphics[scale=1]{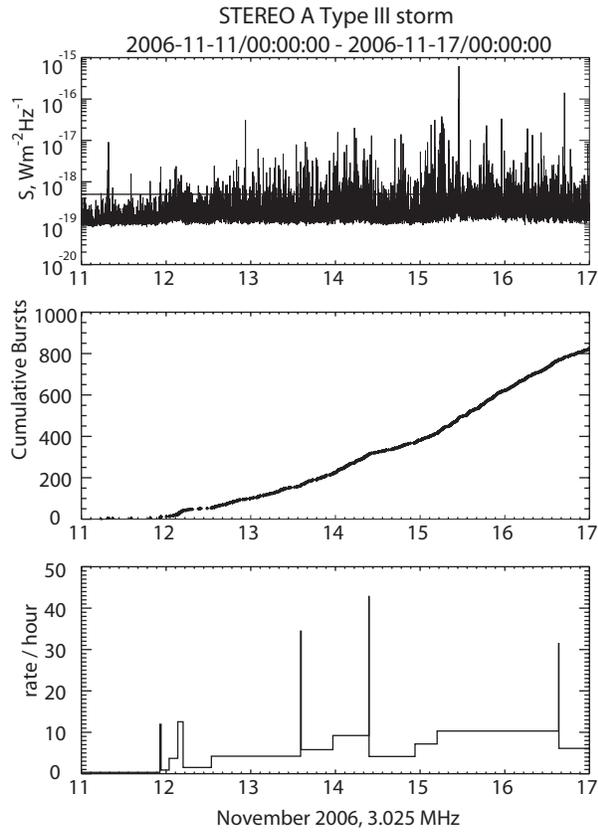}
\caption{\label{fig:fig3}(Top) Time series at 3.025 MHz (middle panel), cumulative number of bursts (bottom panel), and results of the Bayesian blocks decomposition.}
\end{figure}

The top panel of Figure~\ref{fig:fig3} shows the time series at 3.025 MHz, where the horizontal line indicates the threshold used in the analysis. The middle panel of Figure~\ref{fig:fig3} shows the cumulative number of bursts as a function of time. It can be seen that the number of bursts does not increase at a constant rate, rather there are intervals where the number of bursts increases slowly, followed by shorter intervals where the burst rate is enhanced, and the cumulative number of bursts increases more rapidly. To quantify this, a Bayesian block procedure \citep{scar98, whea00} was used to decompose the data into intervals consistent with a constant-rate Poisson process. Since a Poisson process has a waiting-time distribution given by 

\begin{equation}
\label{eq5}
P(\Delta t) = \lambda e^{-\lambda \Delta t}
\end{equation}

\begin{flushleft}
where $\Delta t$ is the interval between events, and $\lambda$ is the rate, a piecewise constant Poisson process made up of $M$ intervals with rates $\lambda_i$ and durations $t_i$ has a WTD given by:
\end{flushleft}

\begin{equation}
\label{eq6}
P(\Delta t) \approx \frac{1}{\bar{\lambda}} \sum_{i=1}^M \frac{t_i}{T}\lambda_i^2 e^{-\lambda_i \Delta t}
\end{equation}

\begin{flushleft}
where $\bar{\lambda}$ is the average burst rate and $T$ is the duration of the observing interval \citep[e.g.][]{whea02}. The lower panel of Figure~\ref{fig:fig3} shows the Bayesian blocks rate decomposition. The Bayesian procedure decided that there were $M$ = 16 intervals with approximately constant rate during the observing period. 
\end{flushleft}

\begin{figure}
\includegraphics[scale=1]{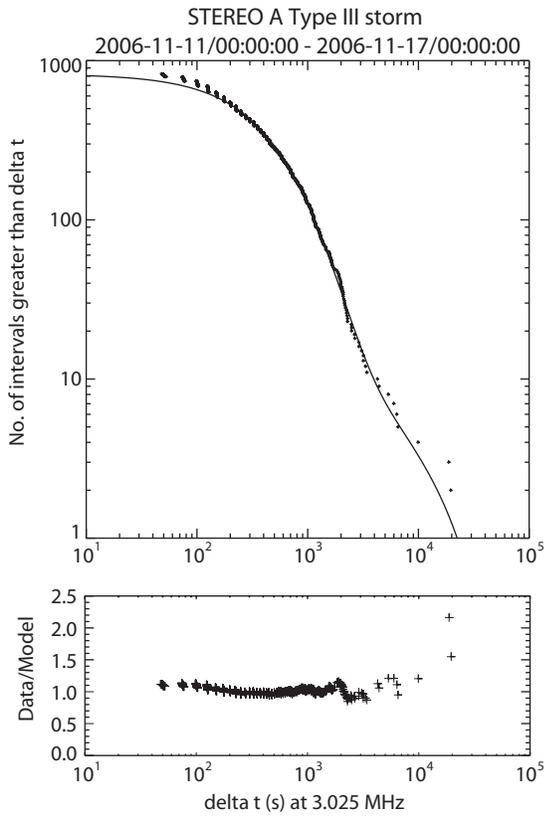}
\caption{\label{fig:fig4}The top panel shows the cumulative waiting time distribution of bursts at 3.025 MHz, together with model based on a piecewise constant Poisson process, with rates determined by the Bayesian blocks procedure. The bottom panel shows the ratio of the data and the model.}
\end{figure}

The top panel of Figure~\ref{fig:fig4} shows the observed WTD as points. The data are again presented cumulatively to avoid binning of the data. The model cumulative WTD is given by:

\begin{equation}
\label{eq7}
P(\Delta t > \Delta t_0) = \int^{\infty}_{\Delta t_0} P( \Delta t)  d\Delta t \approx \frac{1}{\bar{\lambda}} \sum_{i} \frac{t_i}{T}\lambda_i e^{-\lambda_i \Delta t_0}
\end{equation}

\begin{flushleft}
where the rates $\lambda_{\rm{i}}$ and intervals $t_{\rm{i}}$ come from the Bayesian blocks procedure. This model is also shown in the top panel of Figure~\ref{fig:fig4}. The bottom panel of Figure~\ref{fig:fig4} shows the ratio of the data and the model as a function of waiting time. There is good agreement between the data and the model, which strongly suggests that the observed WTD is consistent with a piecewise constant Poisson process whose rate varies as shown in Figure~\ref{fig:fig3}. Similar results were obtained for the different frequencies shown in Table~\ref{table1}.
\end{flushleft}

\section{Discussion}

As mentioned in the introduction, type III emission is thought to arise from non-thermal electron beams on open field lines. Emission may occur at the fundamental or its harmonic; type III bursts whose trajectory intersects the spacecraft show that the emission is at the fundamental mode \citep{lebl98,dulk2000}, but the possibility of harmonic emission perhaps should not be discounted \citep{kell80, boug84b}. Since the local electron plasma frequency depends on density, that is $f_{p} [\rm{Hz}] = 8.98\times10^{3} \times (\it{n}_{\rm{e}} [\rm{cm}^{-3}])^{1/2}$, a density model is required to determine the height of the radio emission. Using the model developed by \citet[]{lebl98}, in this example 2 MHz corresponds to $\sim$ 3 $R_{S}$ and 9 MHz corresponds to $\sim$ 1.8 $R_{S}$. The model is scaled to the solar wind density at 1 AU; between 17 - 20 November 2006 the average solar wind density measured by the ACE spacecraft at 1 AU is approximately 4 $\rm{cm}^{-3}$. If the type III emission is harmonic, then then the local density is reduced, and the emission comes from higher in the corona - between approximately 2.5 and 9 $R_{S}$ for 9 MHz and 2 MHz. 

It is generally accepted that the radio emission arises from the non-linear conversion of Langmuir waves excited at the local plasma frequency by non-thermal electron beams. As the electrons move along the magnetic field, the faster electrons outrun the slower electrons leading to a bump-on-tail distribution and instability \citep{robi00}. The growth of Langmuir waves causes the distribution to relax to marginal stability; in stochastic growth theory, fluctuations in the ambient density are thought to cause fluctuations about marginal stability which allows wave growth without rapid destruction of the electron beam \citep{robi92, robi93}. This suggests that the electron beams are being emitted from closer to the Sun. The more general association of type III storms (open field) and type I storms (closed field) suggests that the storm activity is related to reconfiguration of the boundary between open and closed magnetic fields in the underlying active region. Analysis of Potential Field Source Surface maps \citep{schr03} shows that a small region of open field lines was persistently present adjacent and to the west of the active region during the storm. The interface between the open and closed field regions rotated behind the limb before the active region itself, which may help to explain the observed decline in storm activity prior to the active region rotating past the limb.

In this type III storm, a $S^{-2.1}$ power law was found in the number distribution, corresponding to a $S^{-3.1}$ power law in the number/unit flux density. The power law index was found to be largely independent of frequency. Although the exact relationship between the primary energy release, the energy and properties of the electron beam, and the energy converted to radio emission is not well understood, it is likely that this reflects power-law behavior in the underlying dynamics. In a previous study, \citet{mori07} observed a $S^{-3.6}$ power-law in the number distribution of the storm (`micro') burst population, but this was averaged over several storm intervals and used data binning. In this analysis, maximum likelihood was used, although this technique should be used with care if there are departures from the power-law behavior due to the method of event selection, background subtraction, etc.  It is necessary to study other events to understand whether this difference represents variability from storm to storm, or is due to some other effect. The rate at which bursts are observed is found to change through the storm. Using a Bayesian block decomposition, we find the observed WTD to be consistent with a piecewise-constant Poisson process, which indicates that individual type III bursts occur independently of one another. It is interesting to note that the Bayesian block structure has no correlation with the occurrence of flares in the active region. A GOES C3.3 event occurred on 12 November 2009 at 2155 UT, for example, in the middle of a long Bayesian block.

These two properties are predicted by avalanche models of driven systems. In the solar context, such models have been developed to understand the properties of solar flares \citep{lu1993}. This suggests that in the active region under consideration, magnetic energy and flux, slowly injected from below and being driven e.g. by twisting due to foot point motion, is being released in a series of independent events which result in the emission of electron beams onto open field lines. In particular, the long lifetime of type III storms indicates that the active region can remain in such a dynamically balanced state for a prolonged period of time even as the burst rate slowly changes. This stability was not disturbed by the $\sim$ 21 GOES flares in this time interval from AR 10923, including 5 C-class events before 14 November 2006.

\acknowledgments

This work was supported by NASA and CNES. HSH and SK acknowledge support from NAS grant 5-98033.

{\it Facilities:} \facility{STEREO}.

\clearpage



\clearpage
\begin{deluxetable}{cc}
\tablecaption{Power law index as a function of frequency\label{table1}}
\tablewidth{0pt}
\tablehead{\colhead{Frequency (MHz)} & \colhead{Power Law Index}}
\startdata
3.025 & 2.16 $\pm$ 0.04 \\
4.025 & 2.13 $\pm$ 0.04 \\
5.025 & 2.10 $\pm$ 0.05 \\
6.025 & 2.15 $\pm$ 0.06 \\
7.025 & 2.24 $\pm$ 0.06 \\
8.025 & 2.22 $\pm$ 0.07 \\
9.025 & 2.36 $\pm$ 0.07 \\
10.025 & 2.20 $\pm$ 0.07 \\
\enddata
\end{deluxetable}


\begin{thebibliography}{}
\bibitem[Bai (1993)]{bai93} Bai, T. 1993, \apj, 404, 805
\bibitem[Bale et al. (2008)]{bale08} Bale, S. D., Ullrich, R., Goetz, K., Alster, N., Cecconi, B., Dekkali, M., Lingner, N. R., Macher, W., Manning, R. E., McCauley, J., Monson, S. J., Oswald, T. H., \& Pulupa, M. 2008, \ssr, 136, 529
\bibitem[Bougeret, Fainberg \& Stone (1984a)]{boug84a} Bougeret, J.-L., Fainberg, J., \& Stone, R. G. 1984a, \aap, 136, 255
\bibitem[Bougeret, Fainberg \& Stone (1984b)]{boug84b} Bougeret, J.-L., Fainberg, J., \& Stone, R. G. 1984b, \aap, 141, 17
\bibitem[Bougeret et al. (2008)]{boug08} Bougeret, J.-L., et al. 2008, \ssr, 136, 487
\bibitem[Cane (1979)]{cane79} Cane, H. V. 1979, \mnras, 189, 465
\bibitem[Dulk (2000)]{dulk2000} Dulk, G. A. 2000, Radio Astronomy at Long Wavelengths Geophysical Monograph 119 (AGU Washington DC), 115
\bibitem[Eastwood et al. (2009)]{east09} Eastwood, J. P., Bale, S. D., Maksimovic, M., Zouganelis, I., Goetz, K., Kaiser, M. L., \& Bougeret, J.-L. 2009, Radio Sci., 44, RS4012, doi:10.1029/2009RS004146
\bibitem[Fainberg \& Stone (1970a)]{fain70a} Fainberg, J., \& Stone, R. G. 1970a, \solphys, 15, 222
\bibitem[Fainberg \& Stone (1970b)]{fain70b} Fainberg, J., \& Stone, R. G. 1970b, \solphys, 15, 433
\bibitem[Fainberg \& Stone (1971)]{fain71} Fainberg, J., \& Stone, R. G. 1971, \solphys, 17, 392
\bibitem[Fitzenreiter, Fainberg \& Bundy  (1976)]{fitz76} Fitzenreiter, R. J., Fainberg, J., \& Bundy, R. B. 1976, \solphys, 46, 465
\bibitem[Gopalswamy (2004)]{gopa04} Gopalswamy, N. 2004, \planss, 52, 1399
\bibitem[Kaiser et al. (2008)]{kais08} Kaiser, M. L., Kucera, T. A., Davila, J. M., St. Cyr, O. C., Guhathakurta, M., \& Christian, E. 2008, \ssr, 136, 5
\bibitem[Kayser et al. (1987)]{kays87} Kayser, S. E., Bougeret, J.-L., Fainberg, J., \& Stone, R. G. 1987, \solphys, 109, 107
\bibitem[Kellogg (1980)]{kell80} Kellogg, P. 1980, \apj, 236, 696
\bibitem[Leblanc, Dulk \& Bougeret (1998)]{lebl98} LeBlanc, Y., Dulk, G. A., \& Bougeret, J.-L. 1998, \solphys, 183, 165
\bibitem[Lu et al. (1993)]{lu1993} Lu, E. T., Hamilton, R. J., McTiernan, J. M., \& Bromund, K. R. 1993, \apj, 412, 841
\bibitem[Mercier \& Trottet (1997)]{merc97} Mercier, C. \& Trottet, G. 1997, \apj, 474, L65
\bibitem[Morioka et al. (2007)]{mori07} Morioka, A., Miyoshi, Y., Masuda, S. Tsuchiya, F., Misawa, H., Matsumoto, H., Hashimoto, K., \& Oya, H. 2007, \apj, 657, 567
\bibitem[Reiner et al. (2001)]{rein01} Reiner, M., Kaiser, M. L., Karlicky, M., Jiricka, K. \& Bougeret, J.-L. 2001, \solphys, 204, 123
\bibitem[Robinson (1992)]{robi92} Robinson, P. A. 1992, \solphys, 139, 147
\bibitem[Robinson, Cairns \& Gurnett (1993)]{robi93} Robinson, P. A, Cairns, I. H., \& Gurnett, D. A., \apj, 407, 790
\bibitem[Robinson \& Cairns (2000)]{robi00} Robinson, P. A. \& Cairns, I. H. 2000, Radio Astronomy at Long Wavelengths Geophysical Monograph 119 (AGU Washington DC), 37
\bibitem[Scargle (1998)]{scar98} Scargle, J. 1998, \apj, 504, 405
\bibitem[Schrijver \& DeRosa (2003)]{schr03} Schrijver, C. J. \& DeRosa, M. L. 2003, \solphys, 162, 129 
\bibitem[Wheatland (2000)]{whea00} Wheatland, M. S. 2000, \apj, 536, L109
\bibitem[Wheatland \& Litvinenko (2002)]{whea02} Wheatland, M. S. \& Litvinenko, Y. E. 2002, \solphys, 211, 255
\bibitem[Wheatland (2003)]{whea03} Wheatland, M. S. 2003, \solphys, 214, 361
\bibitem[Wheatland (2004)]{whea04} Wheatland, M. S. 2004, \apj, 609, 1134

\end{thebibliography}
\end{document}